\documentclass[iop]{emulateapj}

\usepackage{amssymb,amsmath}
\usepackage{natbib}
\usepackage{graphicx}
\usepackage{epstopdf,color}
\def\ale{\mathrel{\hbox{\rlap{\hbox{\lower4pt\hbox{$\sim$}}}\hbox{$<$}}}}
\def\age{\mathrel{\hbox{\rlap{\hbox{\lower4pt\hbox{$\sim$}}}\hbox{$>$}}}}

\definecolor{orange}{cmyk}{0,0.4,0.8,0.2}
\definecolor{darkorange}{rgb}{.71,0.21,0.01}
\definecolor{darkgreen}{rgb}{.12,.54,.11}
\definecolor{darkblue}{rgb}{0.1,0.1,0.8}

\usepackage{hyperref}
\hypersetup{pdftex,  % needed for pdflatex
  breaklinks=true,  % so long urls are correctly broken across lines
  colorlinks=true,
  urlcolor=blue,
  linkcolor=darkorange,
  citecolor=darkgreen,
  }

\def\nar{{\rm New~Ast.~Rev.}}        

\def\snfe{SN\,2011fe}
\begin{document}

\title{A Compact Degenerate Primary-Star Progenitor of \snfe}
\author{{\sc Joshua S. Bloom}\altaffilmark{1,2}, {\sc Daniel Kasen}\altaffilmark{3,4,1}, 
        {\sc Ken J. Shen}\altaffilmark{1,3,$\dagger$}, {\sc Peter E. Nugent}\altaffilmark{2,1}, 
        {\sc Nathaniel R. Butler}\altaffilmark{5}, {\sc Melissa L. Graham}\altaffilmark{6,7},
        {\sc D. Andrew Howell}\altaffilmark{6,7}, {\sc Ulrich Kolb}\altaffilmark{8}, 
        {\sc Stefan Holmes}\altaffilmark{8}, {\sc Carole Haswell}\altaffilmark{8},
        {\sc Vadim Burwitz}\altaffilmark{9},{\sc Juan Rodriguez}\altaffilmark{10},
        {\sc Mark Sullivan}\altaffilmark{11}}

\slugcomment{Submitted to ApJL, October 17, 2011}

\shorttitle{A Compact Degenerate Primary-Star Progenitor of \snfe}
\shortauthors{BLOOM ET AL.}

\altaffiltext{1}{Department of Astronomy, University of California, Berkeley, Berkeley  CA, 94720-3411, USA; jbloom@astro.berkeley.edu}

\altaffiltext{2}{Physics Division, Lawrence Berkeley National Laboratory, Berkeley, CA 94720, USA}

\altaffiltext{3}{Nuclear Science Division, Lawrence Berkeley National Laboratory, Berkeley, CA 94720, USA}

\altaffiltext{4}{Department of Physics, University of California, Berkeley, Berkeley  CA, 94720, USA}

\altaffiltext{$\dagger$}{Einstein Fellow}

\altaffiltext{5}{School of Earth and Space Exploration, Arizona State University, P.O.\ Box 871404, Tempe, AZ 85287-1404, USA}

\altaffiltext{6}{Las Cumbres Observatory Global Telescope Network, 6740 Cortona Dr., Suite 102, Goleta, California 93117, USA}

\altaffiltext{7}{Department of Physics, University of California Santa Barbara, Santa Barbara, CA 93106, USA}

\altaffiltext{8}{Department of Physical Sciences, The Open University, Walton Hall, Milton Keynes MK7 6AA, UK}

\altaffiltext{9}{Max-Planck-Institut f\"ur extraterrestrische Physik, Giessenbachstrasse, 85748 Garching, Germany}

\altaffiltext{10}{Observatori Astron\`omic de Mallorca, Cam\'i de l'Observatori, 07144 Costitx, Mallorca, Spain}

\altaffiltext{11}{Department of Physics (Astrophysics), University of Oxford, Keble Road, Oxford OX1 3RH, UK}

\begin{abstract}
While a white dwarf is, from a theoretical perspective, the most plausible primary star in Type Ia supernova (SN Ia), many other candidates have not been formally ruled out. Shock energy deposited in the envelope of any exploding primary contributes to the early SN brightness and, since this radiation energy is degraded by expansion after the explosion, the diffusive luminosity depends on the initial primary radius. We present a new non-detection limit of the nearby SN Ia 2011fe, obtained what appears to be just 4 hours after explosion, allowing us to directly constrain the initial primary radius, $R_p$. Coupled with the non-detection of a quiescent X-ray counterpart and  the inferred synthesized $^{56}$Ni mass, we show that $R_p \ale 0.02 R_\odot$ (a factor of 5 smaller than previously inferred), that the average density of the primary must be $\rho_{p}> 10^4$ gm cm$^{-3}$, and that the effective temperature must be less than a few $\times 10^5$ K. This rules out hydrogen burning main sequence stars and giants. Constructing the helium-burning main sequence and carbon-burning main sequence, we find such objects are also excluded. By process of elimination, we find that only degeneracy-supported compact objects---WDs and neutron stars---are viable as the primary star of \snfe.  With few caveats, we also restrict the companion (secondary) star radius to $R_{\rm c} \ale 0.1~R_\odot$, excluding Roche-Lobe overflowing red giant and main-sequence companions to high significance.
\end{abstract}

\keywords{supernovae: general---supernovae: individual (2011fe)---white dwarfs}

\section{Introduction}

While the nature of the explosion that leads to a Type Ia supernova (SN Ia)---detonation \citep{1986ApJ...301..601W,1994ApJ...423..371W,1995ApJ...452...62L,2007A&A...476.1133F}, deflagration \citep{1976Ap&SS..39L..37N}, or both \citep{1991A&A...245..114K}---and the process that leads to the explosion trigger are not well known, it is commonly assumed that an SN Ia is powered by the explosion of a white dwarf (WD) at a pressure and temperature sufficient to ignite carbon \citep{1982ApJ...253..798N,1984ApJS...54..335I}.  All viable SN Ia models of the {\it progenitor system} include a companion (secondary) star that transfers mass (either steadily or violently) to the WD \citep{1997Sci...276.1378N,2008NewAR..52..381P}.  Single-degenerate channels involve transfer of mass from a giant, main sequence, or helium star \citep{1973ApJ...186.1007W,1982ApJ...253..798N,1992ApJ...397L..87M,1992A&A...262...97V,2004MNRAS.350.1301H}. Double-generate channels involve the merger of two WDs \citep{1984ApJ...277..355W,1984ApJS...54..335I}. 

There is considerable {\it circumstantial} evidence that the primary is a C+O WD \citep{1960ApJ...132..565H}, but until recently (\citealt{thenuge}; see also \citealt{2011arXiv1110.2538B} for a similar analysis) there have been very few direct constraints.  The evidence is (1) neither hydrogen nor helium is  seen in SNe Ia \citep{2007ApJ...670.1275L}, and few astrophysical objects lack these elements,  (2) the elements synthesized in SN Ia are consistent with the fusion chain leading from carbon going up to the iron peak, (3) degenerate objects can result in runaway, explosive burning, (4) the energy gained from burning a WD matches that seen in a SN Ia, (5) simulations of the explosions of CO WD stars are successful at reproducing SN Ia spectra (e.g.,  \citealt{1982ApJ...253..798N}).  However, the discovery of a class of SNe Ia that require a WD mass above the Chandrasekhar limit \citep{2006Natur.443..308H} has caused some to question whether a WD is involved in the explosion after all \citep{2011MNRAS.412.2735T}.  

A normal SN Ia, \snfe, was discovered in the Pinwheel Galaxy (M101) more than two weeks before it hit maximum brightness on 12 Sept 2011 UT \citep{thenuge}. To date, \snfe\ provides some of the best constraints on the progenitor system of an SN Ia: coupled with a well-measured distance modulus to M101 (DM=29.05 $\pm$ 0.23 mag; \citealt{2011ApJ...733..124S}), the non-detection of a quiescent counterpart at optical, infrared, mid-infrared and X-ray wavebands were used to placed strict limits on the nature of the progenitor system.  With optical imaging reaching $\sim$100 times fainter than previous limits, \citet{theli}, \citet{thenuge}, and \citet{2011arXiv1109.2912H} showed that Roche-lobe overflow red giants as the secondary star were excluded; He-star + WD progenitor systems were also largely excluded. \citet{thenuge} placed constraints on double-degenerate models based on the non-detection of early-time emission from shock interaction with the disrupted secondary WD material \citep{shen11}.

Rather than focus on the progenitor system as a whole, in this {\it Letter} we investigate what can be gleaned about the primary star, the body directly responsible for powering the SN. \citet{thenuge} have previously noted that the primary size was small ($R_p \ale 0.1 R_\odot$) based on considerations of the early-time light curve. They concluded, based also on carbon and oxygen observed in the early-time spectra, that a C+O WD was the likely primary. Here, coupling a new, more stringent radius measurement and explicitly discussing mass constraints (based on explosive yield) and temperature constraints (from quiescent non-detection) we narrow the parameter space for the primary even further. Whereas the \citet{thenuge} radius constraint was insufficient to rule out carbon-burning main sequence stars, our results appear to exclude such bodies. By process of elimination, we find that a WD (or a neutron star) are the only allowable primary candidates.

\section{Primary Constraints}
\label{sec:constraints}

\subsection{Mass} As a manifestly ``normal'' SN Ia (spectroscopically and in peak brightness), \snfe\ is expected to show the characteristic decline of the half-life of the radioactive process $^{56}$Co$\rightarrow^{56}$Fe, which in turn suggests at least $\sim0.5 M_\odot$ of synthesized $^{56}$Ni powered the early light curve \citep{1996ApJ...457..500H,1999ApJS..124..503M}. We thus consider $M_{{\rm p,lim}} = 0.5 M_\odot$ as a conservative lower limit to the mass of the primary; this value is below the lower mass limit ($M_{{\rm p,lim}} = 0.7$) for sub-luminous SN Ia models \citep{2010ApJ...714L..52S,2011ApJ...734...38W}. While a Chandrasekhar mass, $M_{\rm ch} = 1.4 M_\odot$ is typically invoked for the primary, there are no stringent upper limits on the primary mass. To accommodate so-called super-Chandrasekhar events, we thus consider a conservative mass  range for the primary is $M_p = 0.5$ to $3 M_\odot$.  

\subsection{Radius}
\label{sec:rad}

\begin{figure}[htbp]
\centerline{\includegraphics[width=3.2in]{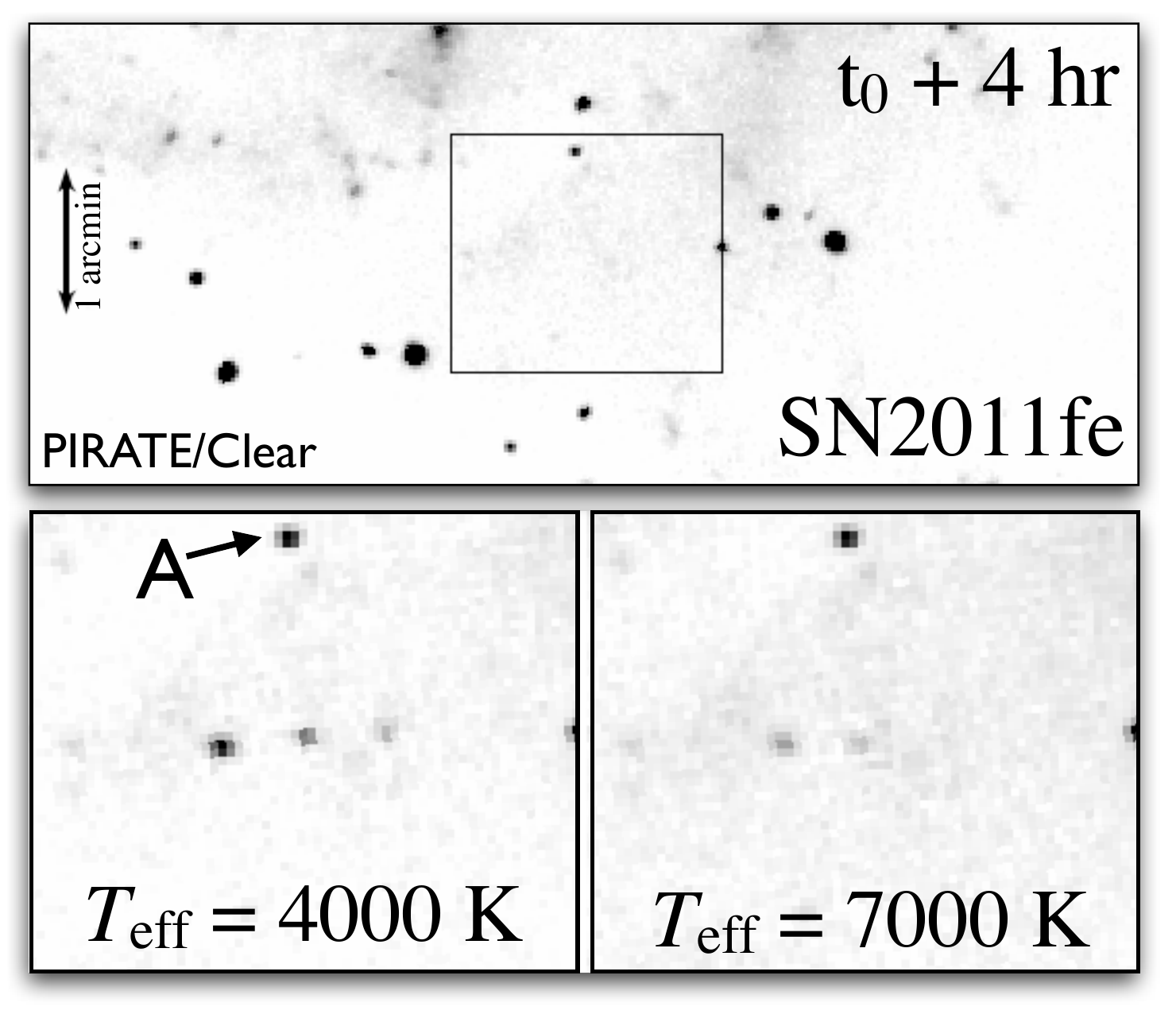}}
\caption{(top) Stacked PIRATE/Clear image obtained starting at $t_0 + 3.38$ hr (mid-point $t_0 + 3.92$ hr) centered on the position of \snfe; no significant flux at the SN position is detected. (bottom, left) Zoomed images near the SN position, showing, from left to right, a fake star with $g=18.5$, 19.0 and 19.5 mag. The star was created using a postage stamp image of an isolated bright star on the stacked image with known SDSS $g$-band magnitude and a color consistent with $T_{\rm eff} = 4000$ K. A shock with $g=19.5$ mag would have been marginally detected. The compact source labelled ``A'' is  SDSS J140306.16+541706.5, with $g=18.20$ and $g - r = 1.16$. (bottom, right) Same as at left, but using a known SDSS star with $T_{\rm eff} = 7000$ K. A shock with $g=19.0$ mag would have been marginally detected.}
\label{fig:im}
\end{figure}

The earliest observations of \snfe\ can be used to constrain the radius of the primary star. The detection at what was determined (from a remarkable several-day $t^2$ behavior of the optical light curve) to be 11 hr post explosion, placed a constraint on the primary to be $R_p \ale 0.1 R_\odot$ \citet{thenuge}. Starting about 7.5 hours before the PTF discovery image, we had fortuitously acquired a series of images of M101, covering the position of \snfe\ for one hour; the data were obtained on PIRATE (Physics Innovations Robotic Astronomical Telescope Explorer) on the The Open University's 0.4m telescope in Mallorca \citep{2011arXiv1108.4196H}\footnote{PIRATE is mainly deployed as part of the SuperWASP consortium for follow-up photometry of exoplanet transit candidates, and for a routine nova monitoring program of M31.}. We analyzed these images and found no significant excess flux at the SN location (Fig.~\ref{fig:im}). Since the filter system was clear we translated the non-detection to a $g$-band magnitude equivalent under the assumption of a range of blackbody temperatures of the shock. For blackbody temperatures in the range 3000--150000 K, we find a robust upper limit of $g = 19.0$ mag at 5 $\sigma$ (at a mean time of $3.92$ hr post explosion). We use this new non-detection to constrain the primary radius under several different shock models.

Initially, the explosion shock wave deposits radiation energy throughout the stellar envelope, which subsequently diffuses out and contributes to the supernova brightness.  Because the radiation energy is lost to adiabatic expansion after the explosion, the luminosity from shock heating depends on the initial stellar radius.   A number of analytical models for the early luminosity have been constructed under the assumption of spherical symmetry \citep[e.g.,][]{chevalier_1992, chevalier_2008, Piro_2010, Kasen_2010, nakar_2010, rabinak_2011a, Rabinak_2011b}. 
The studies differ in their assumptions of e.g., the initial ejecta density and pressure profiles, the nature of the opacity, and in the treatment of radiative diffusion. Nevertheless, the predicted early light curves tend to be quite similar, and have been found to agree with numerical calculations to within a factor of $\sim 2$ \citep{Rabinak_2011b, Kasen_2010}.
For example,  \cite{Rabinak_2011b} find analytic expressions for the luminosity and effective temperature:
\begin{equation}
\begin{split}
L(t) &= 1.2 \times 10^{40}  \frac{R_{10} E_{51}^{0.85}}{ M_c^{0.69} \kappa_{0.2}^{0.85} f_{\rm p}^{0.16} }t_d^{-0.31} ~{\rm ergs~s^{-1}}\\
T_{\rm eff}(t) &= 
4105 ~ \frac{R_{10}^{1/4} E_{51}^{0.016}  M_c^{0.03} \kappa_{0.2}^{0.27}}{ f_{\rm p}^{0.022}}  t_d^{-0.47}~{\rm K}
\end{split}
\label{eq:rabinak}
\end{equation}
where $R_{10}$ is the progenitor radius $R_p/10^{10}~{\rm cm}$, $E_{51}$ is the explosion energy $E/10^{51}~{\rm ergs}$, $M_c$ is the mass in units of $M_{\rm ch}$, and $\kappa_{0.2}$ is the opacity $\kappa/0.2~{\rm cm^2 g^{-1}}$.  These expressions assume a constant opacity, which is appropriate when electron scattering dominates in fully ionized C/O ejecta.  The form factor $f_{\rm p}$ depends on the density profile of the primary star, and has estimated values in the range $0.031$ and $0.13$ \citep{calzavara_2004,Rabinak_2011b}. 

\begin{figure}[htbp]
\centerline{\includegraphics[width=3.5in]{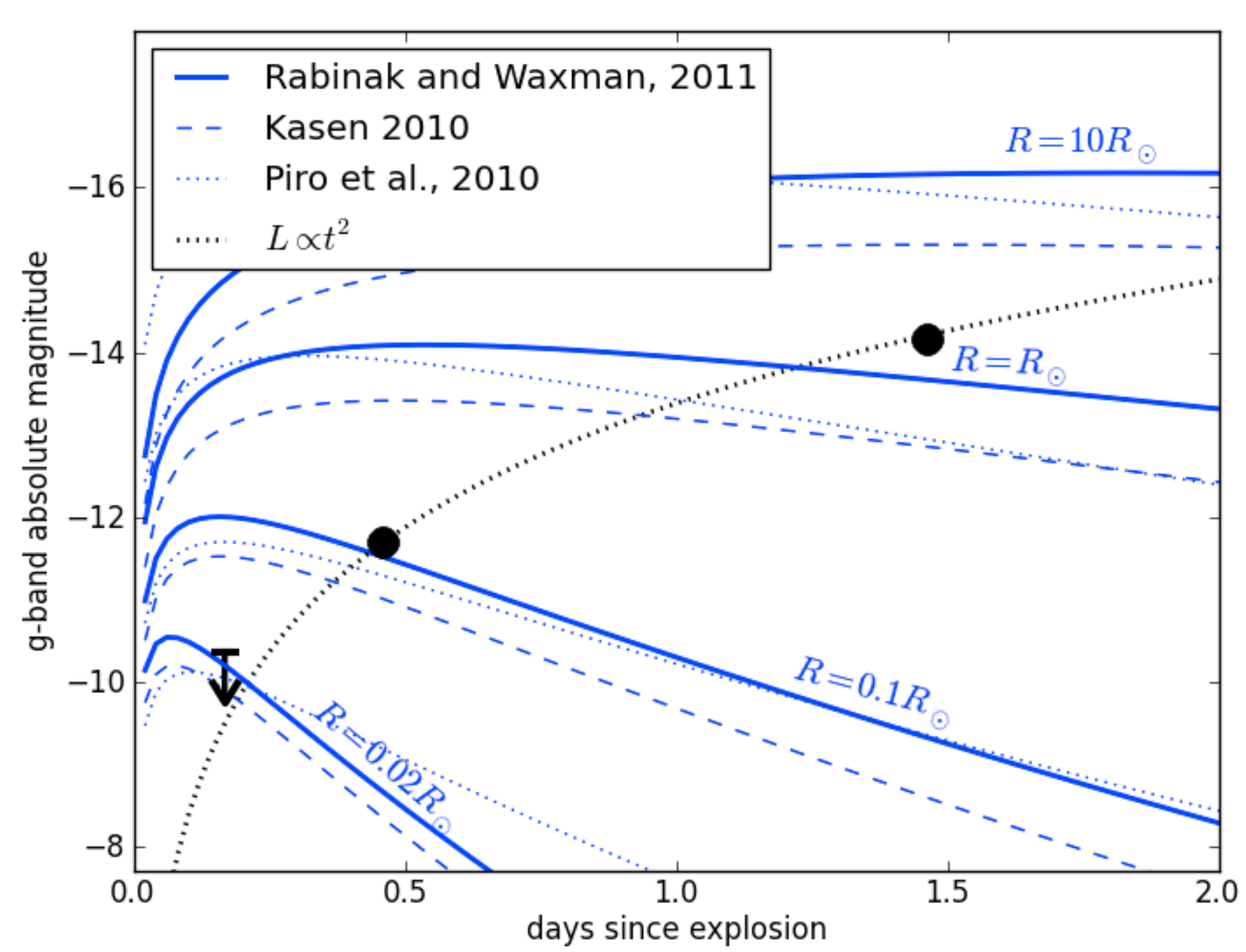}}
\caption{Absolute g-band magnitude versus time since explosion in three theoretical models for the early time evolution of Type Ia SNe. Shown is 4 hr, 5 $\sigma$ non-detection discussed in \S \ref{sec:rad} and the first two detections from PTF \citep{thenuge}. The black line shows the $L\propto t^2$ behavior of the early-time light curve detected in PTF, consistent with the non-detection. For the Kasen (2010) companion interaction model, $R$ denotes the separation distance between the two stars, and the light curve is shown for an observer aligned with the collision axis, which produces the brightest observed luminosity.}
\label{fig:lc}
\end{figure}

We constructed theoretical light curves of the early optical luminosity assuming the spectrum was given by a blackbody with $T_{\rm eff}$, and the flux at the effective wavelength of the filter (given the temperature) was consistent with the non-detection. Figure~\ref{fig:lc} shows the results for a selection of different analytical models, assuming  $E_{51}/M_c = 1$ (constant explosive yield per unit mass), $f_p = 0.05$, $\kappa_{0.2} = 1$ and a variety of values of $R_p$.  The PIRATE observation at 4~hours is the most constraining data point, which limits $R_p \la 0.02 R_\odot$ .  Table~1 summarizes the detailed radius constraints under different assumptions of progenitor mass and under the different models.

The  expressions we have used for the early luminosity hold only under the assumption that radiation energy dominates in the post-shock ejecta.  In fact, the diffusion wave will eventually recede into higher density regions of ejecta where gas pressure dominates.  The luminosity is then expected to drop suddenly; \cite{Rabinak_2011b} show that, for constant opacity, the time of this drop is proportional to $R_p$, which effectively limits the minimal progenitor radius that we are capable of probing.  From their expression for $t_{\rm drop}$ we find this minimal radius to be
\begin{equation}
R_{\rm p, min} \approx 0.013~t_{4 {\rm h}} E_{51}^{-0.66} M_c^{0.56} f_{0.05}^{0.15}~R_\odot
\end{equation}
where $t_{4 {\rm h}} = t/4$~hours.  The  value of $R_{\rm min}$ is just smaller than our limits on $R_p$ determined in Table~1, suggesting that the breakdown of radiation energy domination is not likely to undermine our results.

The early photometry of \snfe\ also tightly constrain the nature of a possible companion star.  The  interaction of the supernova ejecta with a companion star produces emission which depends linearly on the separation distance (Kasen 2010).  This emission will be anisotropic and vary by a factor of $\sim 10$ depending on the orientation.  Assuming the companion star in Roche Lobe overflow, such that its radius is $\ale 1/2$ of the separation distance, and that the observer's viewing angle is unfavorable (such that the light curve is fainter by a factor of 10 from its maximum) our data restricts the companion star radius to $R_{\rm c} \ale 0.1~R_\odot$. Unless the time since explosion for the PIRATE data is vastly underestimated (by $\age$ day), this apparently excludes Roche-Lobe overflowing red giant and main sequence companions to high significance.

{\it Temperature-Radius}: Non-detections of a quiescent counterpart in Hubble Space Telescope (HST) imaging yield a specific luminosity ($L_\nu$) constraint at certain optical frequencies. With the assumption of a spectrum of the primary, these limits can be turned into a limit on the bolometric luminosity ($L$).  \citet{theli}  considered mostly spectra of an unseen secondary, using model input spectra of red giants to derive $L$ constraints. For a high effective temperature primary, here we consider a blackbody as the input spectrum and solve for the bolometric luminosity and effective radius using Stephan-Boltzmann law (see also \citealt{2011arXiv1110.2506L} for a similar analysis). We perform a similar analysis with the Chandra X-ray non-detection, convolving different input blackbody spectra to find a radius limit. At  $10^6$ K, for example, the limits (1,2, and 3 $\sigma$) are $1.2 \times 10^{-3}$ $R_\odot$, $1.5 \times 10^{-3}$ $R_\odot$, and $1.8 \times 10^{-3}$ $R_\odot$.

In Figure \ref{fig:constrain}, we show these primary-star constraints as a function of effective temperature and radius. Primary stars with average density less than $\rho_p$ = 10$^4$ gm cm$^{-3}$ and effective temperatures larger than $10^6$ K (at $\rho_p = 10^{12}$ gm cm$^{-3}$ ) are excluded. 

\section{Comparisons to Primary Candidates}
\label{sec:primary}

\begin{figure}[thbp]
\centerline{\includegraphics[width=4in]{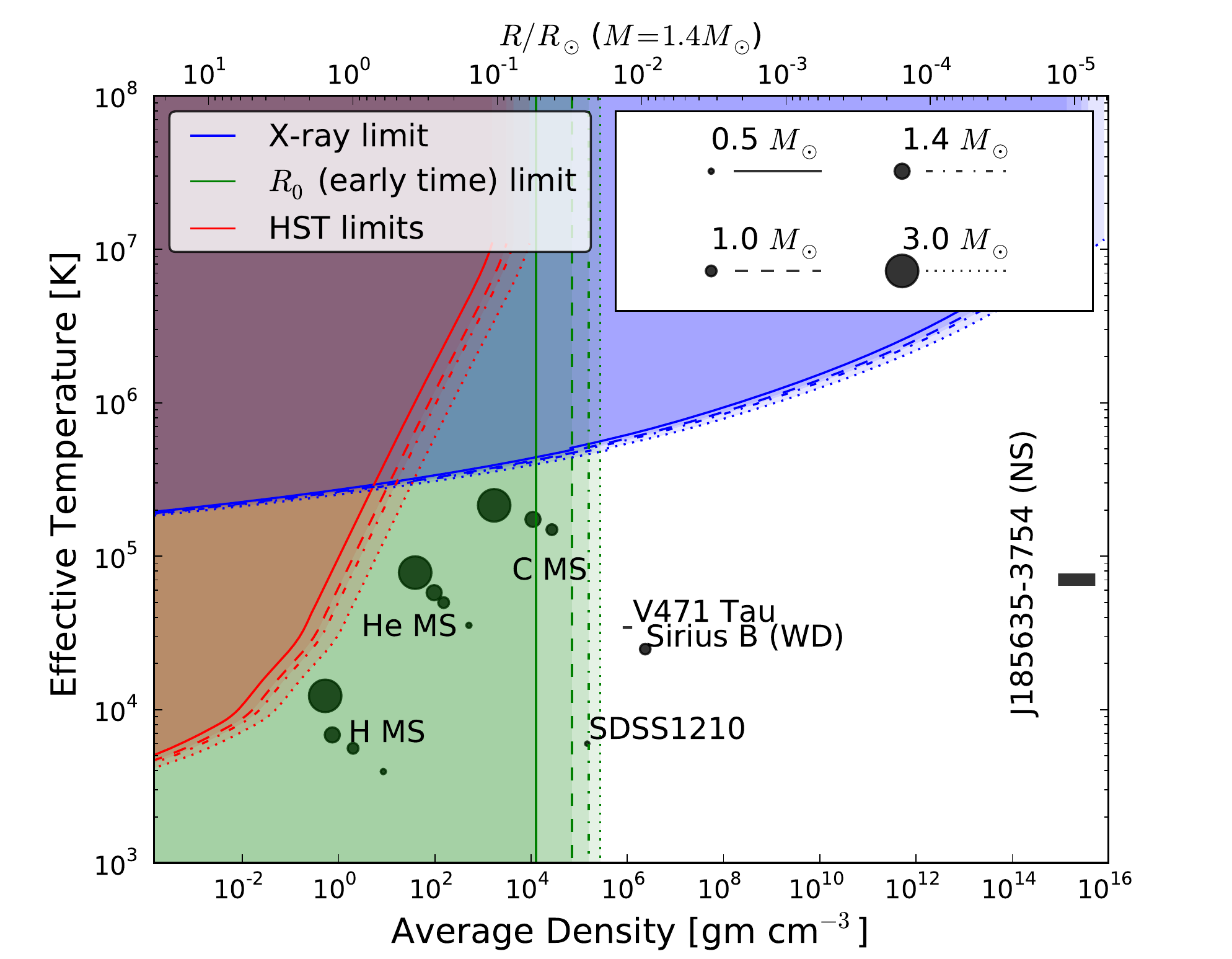}}
\caption{Constraints on mass, effective temperature, radius and average density of the primary star of \snfe. The shaded red region is excluded from non-detection of an optical quiescent counterpart in Hubble Space Telescope (HST) imaging. The shaded green region is excluded from considerations of the non-detection of a shock breakout at early times, taking the least constraining $R_p$ of the three model in Table \ref{tab:cons}. The blue region is excludes by the non detection of a quiescent counterpart in Chandra X-ray imaging. The location of the H, He, and C main sequence is shown, with the symbol size scaled for different primary masses. Several observed WD and NSs are shown. The primary radius in units of $R_\odot$ is shown for $M_p = 1.4 M_\odot$.}
\label{fig:constrain}
\end{figure}

Accepting $0.5 M_\odot$ as a conservative lower limit for the primary mass, low-mass main-sequence stars, brown dwarfs, and planets are not viable. In Figure \ref{fig:constrain}, we show the main sequence of stably H-burning stars with mass 0.5, 1, 1.4, and 3 $M_\odot$. The hydrogen main sequence, shown using solar-metallicity isochrones from \citet{2008A&A...482..883M}, is excluded as the \snfe\ primary. Accepting the radius constraints, giants (not plotted) are also excluded for the primary of \snfe.

We also constructed idealized He- and C-burning main sequence stars with the stellar evolution code MESA \citep{paxt11}.\footnote{http://mesa.sourceforge.net/ (version 3611)}  Uniform composition stars ($X_{\rm 4He}=0.98$ and solar metallicity for the He stars, and $X_{\rm 12C}=X_{\rm 16O}=0.5$ for the C stars) were relaxed until they reached stable non-degenerate equilibrium configurations.  Up-to-date neutrino loss rates \citep{itoh96}, opacities \citep{ir96}, equation of state \citep{ts00b,rn02}, and nuclear reactions \citep{cf88,angu99} were employed, including recently updated triple$-\alpha$ \citep{fynb05}, $\alpha+^{12}$C \citep{kunz02}, and $^{12}$C$+^{12}$C \citep{gasq05} rates. These results are plotted in Figure \ref{fig:constrain} for He star masses of 0.5, 1.0, 1.4, and $3.0 \ M_\odot$, and C star masses of 1.0, 1.4, and $3.0 \ M_\odot$; stable configurations of C stars supported by nuclear burning do not exist below $1.0 \ M_\odot$ \citep{bjs73}.  Comparisons to previous He star calculations (e.g., \citealt{divi65,lecu66}) and C star calculations (e.g., \citealt{sugi68}) match to $\sim10\%$ in the radius and effective temperature. Evidently, He-main sequence stars and C-burning main sequence stars are also excluded as the \snfe\ primary. 

Several WDs in eclipsing binary systems, with measured temperature, radii, and mass, are shown in Figure \ref{fig:constrain} (SDSS1210: \citealt{2011arXiv1109.1171P}; V471 Tau: \citealt{MNR:MNR14378}; Sirius B: \citealt{2005MNRAS.362.1134B}). We also depict isolated neutron star RX J185635$-$3754 ($M \approx M_\odot$), which has strong observed constraints on temperature and density \citep{2002ApJ...564..981P}. These systems are all allowed by our constraints. Note that SDSS 1210 ($M = 0.415 \pm 0.1 M_\odot$; $R = 0.0159 \pm 0.002 R_\odot$; $\rho \approx 1.3-1.6 \times 10^{5}$ gm cm$^{-3}$) is a He WD and has a mass less than our suggested $M_p > 0.5$ constraint. It is also marginally excluded by the \citet{Piro_2010} model.

\begin{deluxetable}{l c c c c}
%\tablewidth{4in}
\tabletypesize{\small}
\tablecaption{Primary Radius and Density Constraints\label{tab:cons}}
\tablehead{
\colhead{$M_{\rm p}$} & 
\colhead{$R_{{\rm p,max}}$\tablenotemark{a}} &  
\colhead{avg.~$\rho_p$} &
\colhead{$L_{\rm shock,max}$} &
\colhead{$T_{\rm shock,max}$} \\
\colhead{$M_\odot$} &
\colhead{$R_\odot$} &
\colhead{gm cm$^{-3}$} &
\colhead{erg s$^{-1}$} &
\colhead{K} 
}
\startdata
\cutinhead{Shock breakout --- Rabinak, Livne, \& Waxman (2011)\tablenotemark{b}}
0.5\ldots	&0.022	&$6.29 \times 10^{4}$	&$4.50 \times 10^{39}$	&6024 \\ 
1.0\ldots	&0.020	&$1.77 \times 10^{5}$	&$4.48 \times 10^{39}$	&6043 \\ 
1.4\ldots	&0.019	&$2.93 \times 10^{5}$	&$4.47 \times 10^{39}$	&6053 \\ 
3.0\ldots	&0.017	&$9.16 \times 10^{5}$	&$4.46 \times 10^{39}$	&6075 \\ 

\cutinhead{Ejecta Heating Secondary --- Kasen (2010)\tablenotemark{c}}
0.5\ldots	&0.038	&$1.26 \times 10^{4}$	&$3.94 \times 10^{39}$	&8048 \\ 
1.0\ldots	&0.027	&$7.05 \times 10^{4}$	&$3.96 \times 10^{39}$	&7388 \\ 
1.4\ldots	&0.023	&$1.58 \times 10^{5}$	&$4.00 \times 10^{39}$	&7103 \\ 
3.0\ldots	&0.017	&$9.29 \times 10^{5}$	&$4.18 \times 10^{39}$	&6531 \\ 

\cutinhead{Shock breakout --- Piro et al. (2010)\tablenotemark{d}}
0.5\ldots	&0.016	&$1.73 \times 10^{5}$	&$5.27 \times 10^{39}$	&12110 \\ 
1.0\ldots	&0.019	&$2.07 \times 10^{5}$	&$5.26 \times 10^{39}$	&12091 \\ 
1.4\ldots	&0.021	&$2.26 \times 10^{5}$	&$5.26 \times 10^{39}$	&12082 \\ 
3.0\ldots	&0.025	&$2.75 \times 10^{5}$	&$5.25 \times 10^{39}$	&12062 
\enddata
\tablenotetext{a}{\scriptsize 5 $\sigma$ limit assuming the 4 hr non-detection (see text) and shock opacity $\kappa$ = 0.2 cm$^2$ g$^{-1}$.}
\tablenotetext{b}{\scriptsize  Assumes $f_p = 0.05$, $E_{\rm 51}/M_c = 1$}
\tablenotetext{c}{\scriptsize  The radius derived is the separation distance and the limit derived is assuming the brightest possible viewing angle. The radius limit comes from the requirement that primary size must be smaller than the semi-major axis of the binary.}
\tablenotetext{d}{\scriptsize  Using their eqns.~35 \& 36 but corrected by a factor of $7^{-4/3}$ ($L$) and  $7^{-1/3}$ ($T_{\rm eff}$) to fix the improper scalings.}
\end{deluxetable}
\section{Summary and Discussion}
\label{sec:disc}

We have placed limits on the average density, effective temperature, and radius of the primary (exploding) star of the Type Ia SN \snfe. We consider the $g=19.0$ mag non-detection as conservative and, assuming the $t^2$ behavior accounted for some of the flux at 4 hours, the flux from the shock inferred from this non-detection would necessarily have been even smaller than that derived from $g=19.0$ mag. In this respect, we take the radius constraint of the primary to be very conservative with the important proviso: if the explosion time was significantly earlier than the time inferred from the $t^2$ fit \citep{thenuge}, the radius constraints are less stringent. In particular, if the PIRATE observations occurred at $t_0 + 28$ hr instead of $t_0 + 4$ hr, then $R_p \ale 0.2 R_\odot$ (still sufficient to rule out H and He main-sequences but not the C MS). Another important caveat is that the radius constraint requires shock heating, naturally expected with a deflagration-detonation
transition \citep{1991A&A...245..114K}\footnote{``Double-detonation'' scenarios \citep{1994ApJ...423..371W} could also lead to heating, but with (slightly) different heating than considered by the models in \S \ref{sec:primary}.}. A pure deflagration of a WD that does not produce a strong shock would not exhibit the early-time behavior of the models presented in Table \ref{tab:cons}; however, pure deflagration is disfavored on nucleosynthetic grounds \citep{1984ApJ...279L..23N,1986A&A...158...17T}. With these caveats aside, this is the most stringent limits on the primary radius and temperature of an SN Ia reported to date.

Clearly the density and temperature in the core of a primary are higher than the reported constrained quantities; and since at high density and high central temperature a star may be supported by pressure other than that associated with fermionic degeneracy pressure, we cannot formally exclude all non-degenerate stars. However, by process of elimination we find that  only compact degenerate objects (WD, NS) are allowed as the explosive primary.  This statement comes from considerations almost orthogonal to the traditional spectral modeling in Type Ia SNe that are invoked to claim WDs as the exploding primary\footnote{As noted in \citet{thenuge}, the early observations of C and O in the spectrum are highly suggestive that the primary was a C+O WD.}. Since the explosive nucleosynthetic yield from phase transition of an NS to a quark-star is expected to be very small ($< 0.1 M_\odot$ in r-process elements) and is unlikely to produce light elements \citep{2007A&A...471..227J}, a NS primary is disfavored.

\acknowledgements{We thank Eliot Quataert, Alex Filippenko, Lars Bilsten, Weidong Li, William Lee, and Philipp Podsiadlowski for helpful discussions.}

%\bibliographystyle{apj}
%\bibliography{ms}

\end{document}